# Nanophotonic enhanced two-photon excited photoluminescence of perovskite quantum dots

*Christiane Becker[‡][*], Sven Burger[¶], Carlo Barth[‡][¶], Phillip Manley[‡][¶], Klaus Jäger[‡][¶], David Eisenhauer[‡], Grit Köppel[‡], Pavel Chabera[†], Junsheng Chen[†], Kaibo Zheng[†], Tönu Pullerits[†]*

[‡] Helmholtz-Zentrum Berlin für Materialien und Energie, Albert-Einstein-Str. 16, 12489 Berlin, Germany

[¶] Zuse Institute Berlin, Takustr. 7, 14195 Berlin, Germany

[†] Department of Chemical Physics and NanoLund, Lund University, P.O. Box 124, 22100 Lund, Sweden

* E-mail: christiane.becker@helmholtz-berlin.de

*Abstract*

All-inorganic CsPbBr$_3$ perovskite colloidal quantum dots have recently emerged as promising material for a variety of optoelectronic applications, among others for multi-photon-pumped lasing. Nevertheless, high irradiance levels are generally required for such multi-photon processes. One strategy to enhance the multi-photon absorption is taking advantage of high local light intensities using photonic nanostructures. Here, we investigate two-photon-excited photoluminescence of CsPbBr$_3$ perovskite quantum dots on a silicon photonic crystal slab. By systematic excitation of optical resonances using a pulsed near-infrared laser beam, we observe an enhancement of two-photon-pumped photoluminescence by more than one order of magnitude when comparing to using a bulk silicon film. Experimental and numerical analyses allow relating these findings to near-field enhancement effects on the nanostructured silicon surface. The results reveal a promising approach for significant decreasing the required irradiance levels for multi-photon processes being of advantage in applications like low-threshold lasing, biomedical imaging, lighting and solar energy.

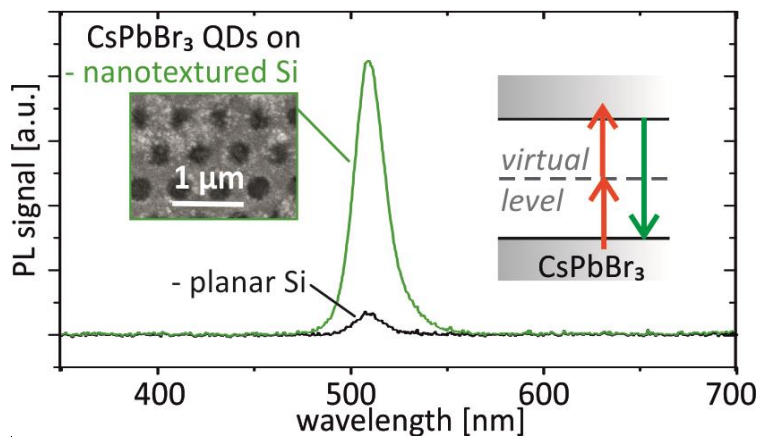

All-inorganic perovskite lead halide semiconductors in the form of colloidal nanocrystals have recently caused a stir as an excellent class of materials for optoelectronic applications [1, 2, 3, 4, 5]. Their advantages range from extremely high photoluminescence efficiencies up to 90%, narrow and tunable emission spectra, facile solution deposition on arbitrary substrates, to the presence of surface-capping ligands for further electronic and optical adjustments. An additional feature of this material family stimulated developments in the field of multi-photon optics: Nanocrystals based on all-inorganic cesium lead bromide ($CsPbBr_3$) perovskite colloidal quantum dots exhibit a large two-photon absorption cross section in the order of $2 \cdot 10^5$ GM [6, 7, 8, 9], inspiring applications on low-threshold multi-photon pumped stimulated emission [6] and lasing [7, 10]. However, multi-photon absorption involving virtual energy levels is generally weak compared to single photon processes and scaling in *n*-th order with excitation intensity, with *n* being the number of absorbed photons [11]. One strategy to enhance the interaction of light with absorbing and photoluminescent species is taking advantage of high local light intensities using metallic (plasmonic) or dielectric (photonic) nanostructures [12]. For instance, photon upconversion by triplet-triplet annihilation or using lanthanide-doped nanophosphors can be enhanced by plasmonic nanostructures [13, 14, 15, 16], partly also in combination with photonic crystal structures [17]. While plasmonic nanostructures have to be carefully designed in order to avoid plasmonic absorption losses in the relevant spectral windows of the device, high refractive index dielectric photonic nanostructures enable reduced dissipative losses and large resonant enhancements [18, 19]. Using two-dimensional photonic crystal slabs, for instance, near-field enhancement effects can cause a significant boost of the (linear) photoluminescence emission from colloidal quantum dots [20, 21]. As a consequence, such two-dimensional nanostructures have become a widely used platform in the field of biosensing and microscopy [22]. Here, the near-field enhancement originates from the excitation of leaky photonic crystal modes spatially overlapping with the emitters, which were directly attached to the nanostructured surface. This experimental configuration - with the propagation direction of light close to the normal of the photonic crystal slab (and not parallel like in many other photonic crystal applications) - very much resembles the current hot topic of dielectric metasurfaces [19].

In this study, we investigate the two-photon excited photoluminescence of $CsPbBr_3$ perovskite quantum dots with 9.4 nm size interacting with the leaky modes of a silicon photonic crystal slab in hexagonal nanohole geometry with a lattice constant of 600 nm. For fabrication of the silicon nanostructures, we apply scalable techniques involving nanoimprint lithography and thin-film growth methods. $CsPbBr_3$ perovskite quantum dots are synthesized by using a method developed by Kovalenko and co-workers [1, 23], and deposited by drop-casting on the silicon nanostructures. By tuning angle of incidence (0° – 50°) and wavelength (900 – 1000 nm) of a pulsed near-infrared laser beam we systematically excite resonance modes of the silicon photonic crystal slab. We measure the two-photon pumped photoluminescence of the $CsPbBr_3$ quantum dots deposited on (i) the silicon photonic crystal slab and (ii) a bulk silicon film. Numerical simulations based on the finite element method relate enhanced photoluminescence in case of the photonic crystal slab to near-field enhancement effects on the nanostructured silicon surface.

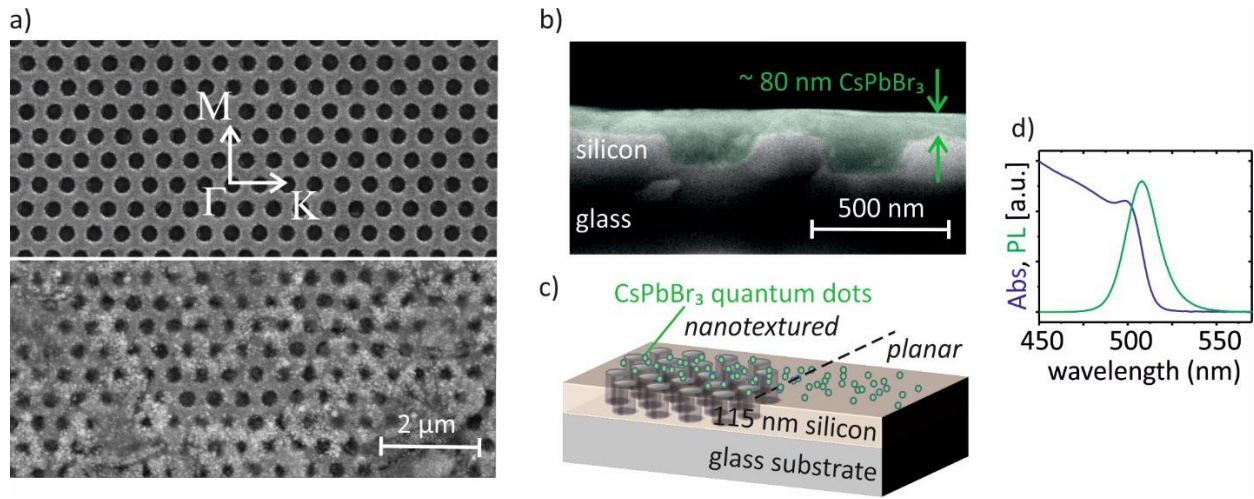

**Figure 1.** Sample description. a) Scanning electron microscope (SEM) image of the silicon nanohole array layer (thickness: 115 nm, hexagonal lattice constant: 600 nm, hole diameter: 365 nm) on glass substrate without (upper part of the image) and with (lower part of the image) drop-casted $CsPbBr_3$ perovskite quantum dot coating. White arrows indicate the high symmetry directions of the hexagonal lattice of the photonic nanostructure. The scale bar applies to both SEM image parts. b) Cross section SEM image with the quantum dot containing layer illustrated in green. c) Schematic of the sample geometry with the quantum dots deposited on nanotextured (left part) and planar (right part) silicon layer as reference (not to scale). d) Absorption (Abs) and photoluminescence (PL) spectra of the $CsPbBr_3$ quantum dots in toluene solution.

For our experiments on nanophotonically enhanced two-photon excited photoluminescence, we use a glass substrate coated by a silicon thin film with 115 nm thickness exhibiting hexagonally arranged cylindrical air holes, 600 nm lattice constant and 365 nm hole diameter radius. Details on the fabrication process based on nanoimprint lithography, physical vapor deposition of silicon, and thermal solid phase crystallization, allowing for nanopatterning of silicon layers on areas up to 5 × 5 cm², are given in the Methods section as well as in references [24, 25, 26]. The upper part of Fig. 1a shows a scanning electron microscope image of the silicon nanohole array. White arrows indicate the two high symmetry directions of the hexagonal lattice, Γ – K and Γ – M. We recently demonstrated numerically and experimentally that such nanopatterned silicon layers form an excellent platform for nanophotonic enhancement of the linear one-photon excited photoluminescence of lead sulfide quantum dots [21]. The underlying mechanism is the systematic excitation of leaky modes of the nanopatterned silicon film by tuning wavelength and angle of incidence of the exciting beam [27]. These leaky modes can exhibit strongly increased electric field energy densities close to the nanostructure surface. Fluorescent species located within the leaky mode volume, i.e. attached to the surface of the nanostructures, show significantly enhanced emission in this case. The aim of the present study is to use such local field enhancement to increase the two-photon pumped photoluminescence of all-inorganic $CsPbBr_3$ perovskite colloidal quantum dots. As two-photon absorption involving a virtual energy level scales quadratically with intensity, we expect the effect of local field enhancement on the photoluminescence being significantly larger than in the case of linear absorption. We coated the nanopatterned silicon layer by 9.4 nm size $CsPbBr_3$ perovskite colloidal quantum dots by drop-casting from toluene solution (see Fig. 1a, lower part). The quantum dots were prepared by using hot-injection method. A Cs-oleate solution was injected into $PbBr_2$ solution at 180°C. Figure 1b depicts the cross section of the sample, pointing out that the quantum dot containing layer fills up the holes of the photonic crystal structure and exhibits a thickness around 80 nm on planar areas of the sample. The variation of the thickness when measuring at different positions of the sample is around ±20 nm. Considering a 1-2 nm capping on the quantum dots this corresponds to about 7 monolayers and a concentration of about 5 to $6 \cdot 10^{12}$ cm$^{-2}$. (Please refer to the Methods Section for details on the concentration calculation.) A schematic drawing of the resulting sample geometry is shown in Fig. 1c (not to scale). A stripe of flat silicon at the edge of the sample serves as reference for the photoluminescence measurements. Hence, the bulk silicon film is placed on the same glass substrate, has the same thickness, and has experienced the same processing sequence during fabrication like the nanopatterned film, only the nanohole pattern is missing. Scanning electron microscopy images reveal that the drop-cast $CsPbBr_3$ quantum dot coating is equivalent on both, nanopatterned and bulk, areas of the silicon film and no significant difference of density and thickness is observed (see Fig. S1 in Supporting Information). Figure 1d exhibits absorption and photoluminescence spectra of the $CsPbBr_3$ perovskite quantum dots in toluene solution. The photoluminescence peak is centered at 508 nm and exhibits a bandwidth of about 20 nm.

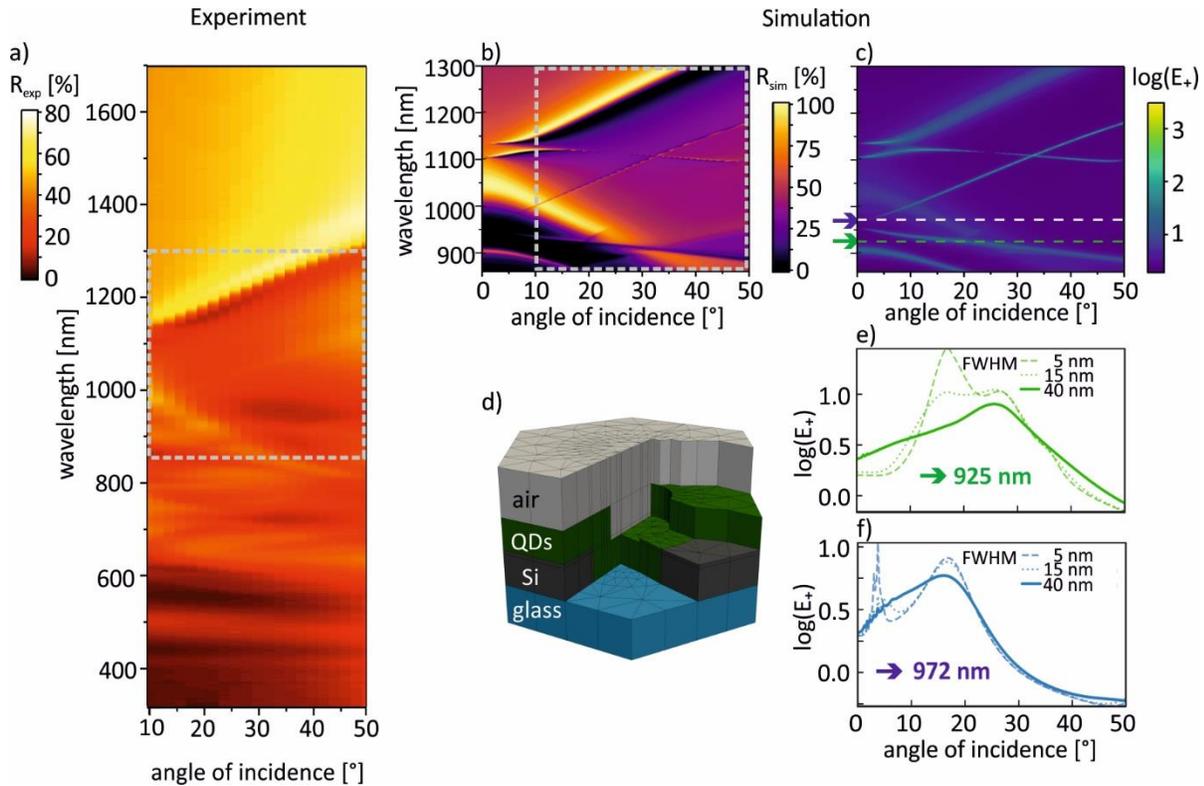

**Figure 2.** Optical properties. **(a)** Experimentally measured angular resolved reflectance $R$ of the silicon nanohole array layer on glass substrate with drop cast $CsPbBr_3$ perovskite quantum dot coating (depicted in Fig. 1a, lower part) using TE-polarized light. The angle of incidence is varied from the surface normal towards the $\Gamma - K$ direction of the hexagonal lattice. The grey box marks the data range selected for analysis by optical simulations. **(b)** Respective

simulated angular resolved reflectance assuming an effective QD-containing layer of 80 nm thickness with effective refractive index $n_{\text{eff}} = 1.7$ covering a silicon photonic crystal slab on glass. Leaky modes appear as resonances in the reflectance spectra. **(c)** Simulated electric field energy enhancement $E_+$ in the CrPbBr$_3$ quantum dot containing layer illustrated on a logarithmic color scale. Arrows and dashed lines indicate the two center wavelengths used for excitation, 925 nm (green) and 972 nm (blue). **(d)** Unit cell of the sample geometry used for the optical simulations (some parts cut away for a better visibility of the cross section). **(e)** and **(f)** Mean field energy enhancement logarithmically displayed assuming Gaussian excitation profiles with center wavelengths 925 nm and 972 nm, respectively, for different FWHM as indicated.

Reflectance spectra are commonly used to investigate photonic band structures in periodically patterned media [27, 28]. Sharp features in such spectra correspond to the excitation of leaky resonance modes. These leaky modes can couple to external light and can exhibit strong near-field enhancement effects close to the surface of the photonic nanostructures [20, 26, 21]. Figure 2a shows the experimentally measured reflectance of the silicon photonic crystal slab in hexagonal nanohole geometry with 600 nm lattice constant, coated by a CsPbBr$_3$ quantum dot layer as depicted in the lower part of Fig. 1a, as function of wavelength and angle of incidence. The incident light was linearly polarized normally to the plane of incidence (TE-polarized) and the angle was tilted with respect to the Γ-K direction of the hexagonal lattice up to 50°. In this limited range of incident angles, we do not expect strongly differing extensions of modes out of the photonic crystal plane for TE and TM modes. (The latter can have a larger out-of-plane extension particularly at grazing incidence.) Resonant features in reflectance changing their spectral position for varying incident angle are clearly visible at wavelengths larger than approximately 600 nm. While in near infrared the resonant features are well distinguishable, the density of features becomes higher towards shorter wavelengths. This can be explained by the high density of modes in the photonic band structure at higher energies (i.e. shorter wavelengths). For wavelengths even shorter than 600 nm the sharp angular dependent resonant features vanish and only a broad angular-independent feature remains. As the absorption coefficient of the silicon is already quite high in that spectral region the penetration depth of the light becomes small. For 508 nm, the wavelength of quantum dot emission, the penetration depths is below 1 μm and hence similar to the photonic crystal periodicity. Coherent effects involving many photonic crystal periods are therefore not expected, explaining the missing resonances in the reflectance spectrum. Figure 2b depicts the respective numerical results calculated using a time-harmonic finite-element Maxwell solver (JCMsuite [29]) with plane wave excitation corresponding to the direction of incidence, wavelength and polarization. We numerically describe the sample geometry as shown in Fig. 2d by a hexagonal unit cell consisting of a glass subspace (blue), a silicon nanohole-layer (dark grey), which is conformally covered with a CsPbBr$_3$ quantum dot coating (green) with 80 nm thickness and with effective refractive index $n_{eff}$ = 1.7. The effective refractive index was estimated by porosity analysis of scanning electron microscopic images, the extrapolated refractive index of bulk CsPbBr$_3$ $n_{CsPbBr}$ ~ 2.2 [30] at a wavelength of 950 nm. The effective refractive index $n_{eff}$ was further refined by comparison of experimental and numerical angular resolved reflectance maps, Fig.2a versus Fig. 2b, for different values of $n_{eff}$. (Please see Fig. S2 in the Supporting Information for further simulated reflectance maps assuming different effective refractive indices.) We observe broad resonances with high reflectance in both, experiment and simulation, for instance, the resonance progressing from λ ~ 1050 nm for ϑ = 0° towards λ ~ 860 nm for ϑ = 40°, and the resonance progressing from λ ~ 1140 nm for ϑ = 0° towards λ ~ 1350 nm for ϑ = 50°. The sharp features clearly visible in the simulations, are less pronounced in the experimental data but still discernible and found at similar positions as in the simulations. We hence conclude that the experimental-numerical agreement justifies qualitative numerical predictions. In order to estimate the local near-fields, which are expected to affect the drop-cast CsPbBr$_3$ quantum dots, we defined the quantity $E_+$ describing the local electric field energy enhancement in the vicinity of the silicon nanostructure (Fig. 2c). $E_+$ is determined by integrating the electric field energy density distribution $u(\mathbf{r})$ over the volume of the CsPbBr$_3$ quantum dot coating. The electric field energy density distribution is given by $u(\mathbf{r}) = \frac{1}{4}\mathbf{D}(\mathbf{r})\mathbf{E}(\mathbf{r})$, where $\mathbf{D}(\mathbf{r})$ is the dielectric displacement field and $\mathbf{E}(\mathbf{r})$ the electric field. Subsequently, the field energy is normalized to the energy of the incident plane wave in the same volume $V$ in a bulk medium $U_{pw} = \frac{1}{4}\epsilon_0\epsilon_r|\mathbf{E}|^2 V$, with $\epsilon_r = n_{\text{eff}}^2$. The value of $E_+$ can therefore be regarded as quantity for the linear intensity enhancement affecting the drop-cast CsPbBr$_3$ quantum dots located on the silicon nanostructure.

Photonic crystal structures can affect the emission of quantum dots by several mechanisms [31, 32]: (i) absorption enhancement of the exciting light, (ii) out-of-plane extraction enhancement of the emitted light, i.e. distinct directions with enhanced light emission [33, 3], an effect often discussed in the context of light emitting diodes [3, 3], and (iii) changes of the spontaneous emission rate (Purcell effect). In our experiment design, effect (ii) and (iii) are expected to play a minor role because of the above-mentioned high absorption of silicon at the emission wavelength (508 nm). The low penetration depth of light suppresses photonic band structure effects. Mechanism (i), however, is relevant in our experiments as excitation takes place at near infrared wavelengths with low absorption in the silicon and a penetration depth much larger than the photonic crystal dimensions ($d_{\text{Si},\lambda=900\text{nm}} \sim 50$ µm). For more details on ruling out emission/extraction effects please also see Fig. S3 in the Supporting Information showing that the local field energy enhancement $E_+$ does not reach values significantly larger than 1 in the range of the emission wavelength (508 nm). For our experiments, we regard the excitation regime ranging from about 900 nm to 1000 nm as most significant, as for wavelengths larger than 1000 nm the perovskite quantum dots do not evince two-photon-absorption and for wavelengths below 900 nm the absorption of the crystalline silicon becomes noticeable such that leaky mode induced effects like near-field enhancements start decreasing [26]. Therefore, the resonances crossing the excitation regime ranging from about 900 nm to 1000 nm are most relevant for our experiments, e.g. the above mentioned broad resonance starting at a wavelength of 1050 nm at normal incidence and proceeding to smaller wavelengths for larger angles. Colored arrows and dashed lines indicate the two excitation center wavelengths chosen in this study (925 nm and 972 nm). At a wavelength of 925 nm a sharp and a broad resonance appear at an incident angle between 20° and 30°. For a wavelength of 972 nm also a sharp (~5°) and a broad (~18°) resonance is crossed. In the resolution of the simulations ($\Delta\lambda$ = 1 nm and $\Delta°$ = 0.25°) this would correspond to $E_+$ enhancement factors up to 1000 for the sharp and up to 10 for broad resonances, respectively. However, in our experiments the excitation laser has a broad spectrum with FWHM up to 40 nm such that the interaction with sharper leaky mode resonances occurring in the same wavelength frame will be significantly diminished. In order to account for a realistic laser excitation spectrum, Fig. 2e and 2f depict *averaged* cuts through the $E_+$ field energy enhancement map assuming Gaussian excitation profiles with different FWHM (as indicated) at center wavelengths 925 nm and 972 nm, respectively. While for the narrow excitation spectra the sharp photonic crystal resonances can still be resolved, at FWHM = 40 nm the broad resonances dominate the enhancement. For such a broad excitation profile the maximum achievable field energy enhancement $E_+$ amounts to around 7.5 and 5.8 for a center wavelength of 925 nm and 972 nm, respectively. Assuming a quadratic intensity dependence of two-photon excited photoluminescence, enhancement factors in the order of 33 to 56 seem to be possible. The photonic crystal design hence targeted on the existence of a well-distinguishable photonic crystal mode in the wavelength range between 900 nm and 1000 nm with spectral width comparable to the width of the excitation laser spectrum. Naturally, the spectral position of photonic crystal resonances can be adjusted by changing the geometrical parameters like period, hole diameter and layer thickness. In our fabrication procedure the change of the first two parameters (period, hole diameter) is quite elaborate as new master structures have to be produced before replication by nanoimprint lithography. However, the parameter "layer thickness" is easily accessible in our process sequence. Therefore, we optimized the photonic crystal structures by tuning the silicon layer thickness. It turned out that a silicon photonic crystal slab with period 600 nm, hole diameter of 365 nm and a layer thickness below 130 nm excellently meets the design objective. For more details on photonic crystal design please see Fig. S4 in the Supporting Information.

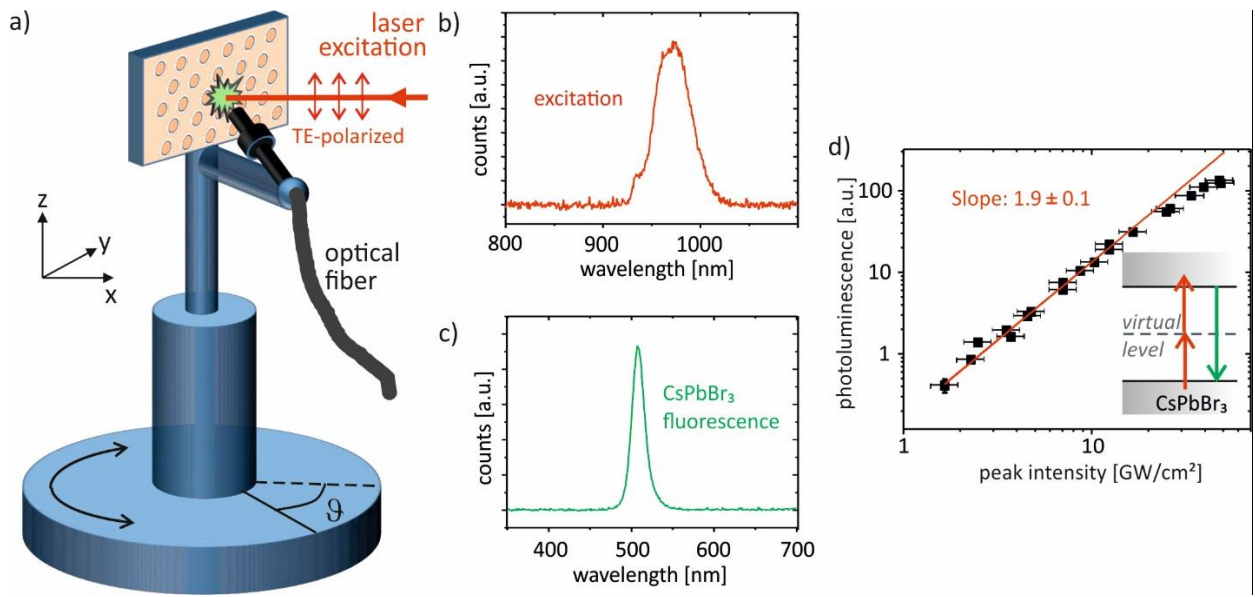

**Figure 3.** Experimental details of the angular resolved two-photon absorption experiment. **(a)** Schematic of the pivoted sample holder, enabling excitation by two-photon absorption at tunable incident angle $\vartheta$. During rotation, the optical fiber collecting the photoluminescence is fixed with respect to the measurement spot, always facing the sample surface at normal incidence in *xy*-plane and at 45° in relation to the vertical *z*-direction. **(b)** Example excitation spectrum of the incident laser beam. **(c)** Example CsPbBr$_3$ photoluminescence spectrum. **(d)** Photoluminescence intensity of CsPbBr$_3$ quantum dots on nanostructured silicon as a function of excitation intensity for an excitation at normal incidence and at $\lambda_{ext}$ = 925 nm. The inset shows the energy level diagram of excitation and photoluminescence of the two-photon absorption experiment.

For controlled excitation of leaky modes of the silicon photonic crystal slab we built a pivotable sample holder as sketched in Fig. 3a. The sample holder allows to adjust the angle $\vartheta$ between sample normal and incident light beam from -20° to +50°. For excitation, we use pulsed, TE-polarized laser light tunable in the range from 900 nm to 1000 nm with a pulse length of around 80 fs and 1 kHz repetition rate. The excitation light is generated by using the second harmonic of the idler of an optical parametric amplifier (Topas, Light Conversion) which is pumped by a regenerative amplifier (Spitfire XP Pro) at 800 nm, 45 fs pulse length and 1 kHz repetition rate. The regenerative amplifier is seeded by a femtosecond oscillator (Tsunami, both Spectra Physics). An example spectrum of the excitation beam from the optical parametric amplifier is shown in Fig. 3b, here with a center wavelength of 972 nm and FWHM of 43 nm. Residual light at lower wavelengths from the optical parametric amplifier was suppressed by using appropriate filters. The excitation laser beam was loosely focused on the sample surface having there a spot diameter of around 500 µm. Therefore, local inhomogeneities on micrometer scale of the silicon photonic nanostructures or the quantum dot distribution as observed in SEM images are effectively averaged out. The drop cast quantum dot coating also exhibits an inhomogeneity on several millimeter scale leading to a changing photoluminescence intensity when moving the sample by several millimeters. Therefore, we adjusted the position of the sample surface carefully such that the spot lies exactly on the rotation axis. This measure ensures a fixed spot position on the sample when rotating the sample over a wide range of rotation angles $\vartheta$, only causing an enlargement of the spot size by $1/\cos\vartheta$, which is well below one millimeter. The optical fiber collecting the photoluminescence of the perovskite quantum dots was attached at a fixed position with respect to the light spot always facing the sample surface at normal incidence in *xy*-plane and at 45° in relation to the vertical *z*-direction (see Fig. 3a) even if the sample is rotated. As circumstantiated above and seen in Fig. 2a and Fig. S3, no extraction enhancement effect, i.e. distinct directions with enhanced light emission, is expected in the wavelength regime of photoluminescence around 508 nm. Therefore, we regard the photoluminescence emission of the CsPbBr$_3$ quantum dots on planar and nanotextured silicon films as isotropic in the half space above the silicon surface. Figure 3c shows an example spectrum of the light collected from CsPbBr$_3$ quantum dots on a nanotextured silicon film after excitation at 925 nm with a peak intensity of $(17 \pm 3)$ GW/cm$^2$. A detailed description of the excitation fluence determination is given in the Methods Section. The photoluminescence peak arising from the CsPbBr$_3$ quantum dots is clearly visible at center wavelength 508 nm with a FWHM of about 20 nm. In order to determine the intensity regime where the photoluminescence signal from the perovskite quantum dots predominantly arises from two-photon-absorption of the excitation laser beam we analyzed the intensity dependence of luminescence. Figure 3d shows the photoluminescence intensity as a function of the excitation laser intensity ($\lambda_{exc}$ = 925 nm) on a log-log scale. For peak intensities below 17 GW/cm$^2$ the slope of the curve is close to 2, i.e. we observe a nearly quadratic intensity dependency as expected for two-photon-absorption processes. The inset illustrates the respective energy level diagram. For higher excitation intensities, however, the photoluminescence increases sub-quadratically with intensity. This is an indication that Auger recombination processes start playing a role. Details on the fitting procedure and setting the upper intensity threshold are given in Fig. S5 in the Supporting Information. In order to rule out degradation of the perovskite quantum dots during the measurements we further performed stability stress tests at high irradiance levels: At very high intensities around 140 GW/cm$^2$ the perovskite quantum dot coating was burned. However, a long-term illumination with an intensity of 58 GW/cm$^2$ yielded no significant degradation of photoluminescence intensity within 12 minutes. We observed no degradation of photoluminescence intensity on a time scale of one day. However, after several weeks (samples stored in air ambient without laser illumination) photoluminescence was slightly decreased. No change of spectral position and width of the photoluminescence peak was observed, neither after several minute illumination (58 GW/cm$^2$) nor after several weeks of storage in air ambient. Based on these stability tests we defined the experimental conditions in order to study the two-photon pumped photoluminescence of CsPbBr$_3$ quantum dots as follows: peak intensities are restricted to below 17 GW/cm$^2$ in order to rule out Auger recombination. Measurements have to be finalized within few days after the quantum dot coating has been drop cast on the silicon photonic nanostructure in order to rule out long-term degradation in air ambient, and series of measurements belonging together should be finalized in the order of several minutes.

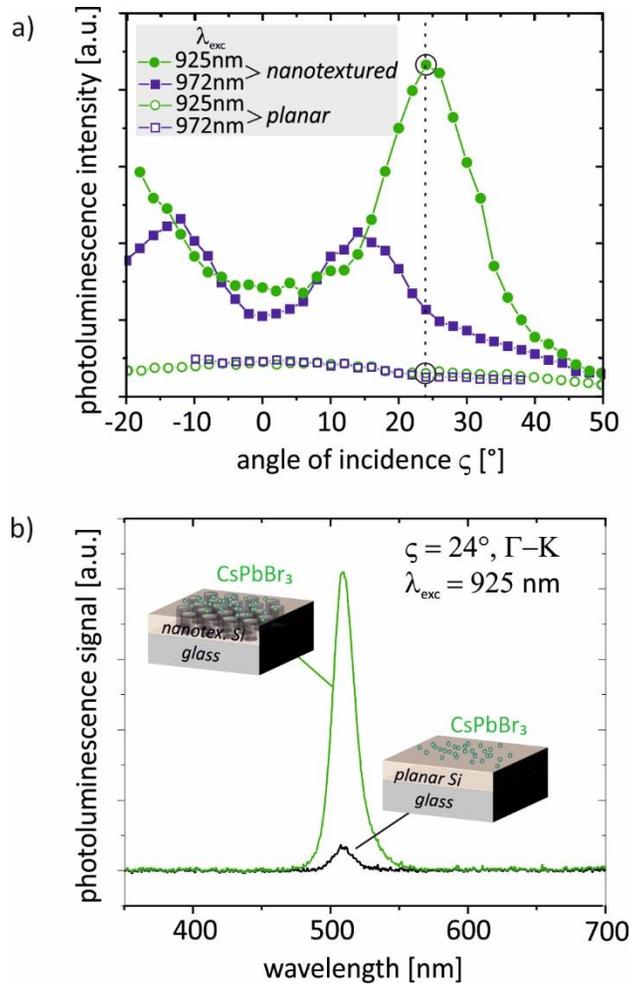

**Figure 4.** Photoluminescence of CsPbBr$_3$ quantum dots with 9.4 nm diameter drop-cast from solution on nanotextured and planar silicon films after excitation by two-photon absorption at intensity $(17 \pm 3)$ GW/cm$^2$ with TE-polarized light. **(a)** Photoluminescence intensity, determined by a Gaussian fit of the photoluminescence peak centered at around 508 nm, as function of the incident angle in $\Gamma - K$ direction of the photonic crystal for excitation at 925 nm (green) and 972 nm (blue), respectively. The excited CsPbBr$_3$ quantum dots are either located on a nanotextured silicon layer (filled symbols) or on a non-textured planar silicon layer (open symbols). In order to ensure the homogeneity of the CsPbBr$_3$ quantum dot coating on both surface morphologies, both measurement areas are located only few millimeters apart from each other. **(b)** Examples of photoluminescence spectra for excitation at $\lambda_{exc} = 925$ nm and an incident angle of $\vartheta = 24°$ in $\Gamma - K$ direction (indicated in part (a) by a dashed line and circles) either with the excitation beam hitting CsPbBr$_3$ quantum dots located on a nanotextured area (green curve) or a planar area (black curve) of the silicon layer.

We measured the intensity of the CsPbBr$_3$ quantum dot photoluminescence for a fixed excitation fluence corresponding to an intensity of $(17 \pm 3)$ GW/cm$^2$ impinging the sample at normal incidence. The incident angle on the sample was varied from -20° to 50° with respect to the Γ-K direction of the hexagonally nanopatterned silicon film, with 0° corresponding to normal incidence. Figure 4a summarizes the results achieved for two different excitation wavelengths, 925 nm (green) and 972 nm (blue), with FWHM of 39 nm and 43 nm, respectively. The two-photon excited photoluminescence of CsPbBr$_3$ quantum dots located on a planar silicon film (open symbols) as well as on the nanopatterned silicon film (filled symbols) is shown. The peaks occur symmetrically with respect to normal incidence at 0°, which is a strong indication for the involvement of leaky modes of the nanopatterned silicon film. A resonance occurring at a particular angle $+X°$ is expected to be also present at an incident angle of $-X°$ due to the mirror symmetry of the hexagonal photonic lattice. The incident angle dependent photoluminescence of the quantum dots on the bulk planar film (open symbols) exhibits a maximum at normal incidence and declines towards larger incident angles in positive as well as negative direction. This can be explained by a reduction of intensity on the sample surface. By tilting the sample by an angle of $\vartheta$, the spot area on the sample surface enlarges by a factor of $1/\cos\vartheta$. As a result, the intensity decreases by a factor of $\cos\vartheta$. In the case of two-photon absorption and a quadratic intensity dependence we expect a decrease of the photoluminescence signal by a factor of $(\cos\vartheta)^2$. Indeed, the measured photoluminescence intensity curves using a planar silicon film as substrate exhibit a $(\cos\vartheta)^2$–characteristic (see Fig. S6 in the Supporting material). When comparing the curves using a planar (open symbols) and a nanopatterned (filled symbols) silicon film as substrate the strongly differing photoluminescence intensity is immediately obvious. The use of a nanopatterned substrate significantly increases the photoluminescence intensity. Furthermore, the photoluminescence intensity curves exhibit characteristic resonances, which are symmetric with respect to normal incidence. For an excitation at 972 nm the maximum photoluminescence signal is observed for incident angles of ±14°. If the quantum dots are excited at 925 nm the maximum photoluminescence occurs at an incident angle of +24°. Figure 4b shows the example photoluminescence spectra of CsPbBr$_3$ quantum dots located on a nanopatterned silicon film (green curve) and on a planar silicon film (black curve) for an excitation at 925 nm at an incident angle of +24°. In this case, we observe an enhancement of two-photon-pumped photoluminescence by a factor of 15 by using a nanopatterned instead of a planar substrate.

In order to explain the experimentally measured photoluminescence enhancements by placing the quantum dots onto a periodically nanostructured surface instead of using a planar substrate, we use the optical simulations as shown in Fig. 2. The two-photon-absorption experiment shown in Fig. 4a can qualitatively be compared to the averaged cut through the wavelength- and angle-dependent local field energy enhancement $E_+$ map at a certain excitation wavelength with large FWHM (Fig. 2e and 2f). In experiment as well as in simulations, the excitation wavelength is kept fixed, while the angle of incidence is varied. As $E_+$ is a linear measure for the intensity enhancement and the two-photon-absorption pumped photoluminescence depends quadratically on the intensity, absolute enhancement values are not comparable. However, the excitation conditions in terms of wavelength and incident angle for maximum photoluminescence can be predicted. The green circles shown in Fig. 4a depict the photoluminescence intensity at 925 nm excitation, which corresponds to the cut through the field energy enhancement map shown in Fig. 2e. Both, experimentally and numerically, a single peak is observed occurring at an incident angle of 24° (experiment) and 26° (simulation), respectively. Further, the blue squares shown in Fig. 4a, depicting the photoluminescence intensity at excitation with center wavelength 972 nm, qualitatively corresponds to the simulated cut through the electric field energy enhancement map shown in Fig. 2f. Here, the highest measured photoluminescence occurs at an incident angle of 14° while the simulated maximum near-field enhancement is calculated for 16°. Qualitative agreement is hence excellent and excitation conditions (angle of incidence and wavelength) yielding maximum two-photon-pumped photoluminescence can be effectively predicted by our numerical model. Regarding quantitative values, the experimentally measured enhancement factors are lower than the numerical prediction (15 versus 56) but in the same order of magnitude. Summing up, the interaction with leaky photonic modes of a two-dimensional photonic crystal slab can easily enhance two-photon pumped photoluminescence of perovskite quantum dots by more than one order of magnitude.

Photoluminescence intensities could eventually be further increased by specifically tuning of several experimental parameters: (1) A better match of the spectral widths of exciting beam and the photonic leaky mode, e.g. by the use of spectral filters or laser pulses with longer pulse duration and hence a narrower spectrum. (2) Choosing a resonance with strong near field enhancement and optimized spatial overlap of the 3D photonic mode distribution with the $CsPbBr_3$ quantum dots. Due to the huge parameter space and the high-dimensional output such an optimization task is challenging. A part of the authors recently introduced a machine learning method based on clustering paving the way for such a qualitative design optimization [36]. (3) The use of tailored defect structures in the periodic photonic nanostructure potentially yielding extremely large intensity enhancements [28].

The simplicity and the robustness of the approach makes it an appealing option in many fields of applications. For instance, two-photon absorption processes play a crucial role in advanced biomedical photonics. Photon-upconversion, lanthanide-doped nanoparticles that emit short-wavelength light under near-infrared excitation enable nearly total elimination of autofluorescence and light scattering of the surrounding matrix. Avoidance of this optical background permits a potentially unprecedented sensitivity [3, 3, 3]. Decreasing the required illumination intensities by the use of nanophotonic structures as described in the present paper might enhance the resolution even further. Additional potential application fields involve low-threshold lasing, bioimaging and biosensing. The choice of an unsophisticated technological platform enabling large-area fabrication opens further possibilities in the solar energy sector. Multi-photon processes enable the shaping of the solar spectrum by up- and down-conversion paving the way to overcome the well-known limit of single-junction photovoltaic devices [40].

In conclusion, the two-photon pumped photoluminescence yield of perovskite quantum dots could be enhanced by more than one order of magnitude by depositing them on a nanostructured silicon film compared to the use of a planar silicon film only. Optical simulations agree well with the experiments and can explain the observed enhanced photoluminescence by near-field enhancement effects of photonic leaky modes. The simple technological platform and robustness of the approach provide opportunities in many fields of applications involving bio imaging and sensing, low-threshold lasing and solar energy.

## Methods

*Synthesis of CsPbB$_3$ quantum dots*

CsPbBr$_3$ colloidal nanocrystals (NCs) were prepared by using a method developed by Protesescu et al. [1, 23] 0.814 g Cs$_2$CO$_3$ (Sigma-Aldrich, 99%) were mixed with 40 mL 1-octadecene (ODE, Sigma-Aldrich, 90%) and 2.5 mL oleic acid (OA, Sigma-Aldrich, 90%), kept at 120 °C and degassed for 1 hour. The mixture was heated up to 150 °C for 30 min under N$_2$ atmosphere. Obtained Cs-oleate was kept in a glove box and heated up to 100 °C before using. Next, 0.0689 g PbBr$_2$ (Sigma-Aldrich, 99.999%) and 10 mL ODE were degassed for 1 hour at 120 °C, afterwards 0.5 mL dry oleylamine (OAm, Sigma-Aldrich, 80−90%) and 0.5 mL OA were added and heated up to 120 °C under N$_2$ atmosphere. The temperature was increased to 180 °C, then 0.4 mL Cs-oleate solution was rapidly injected. After injection, the mixted solution was immediately cooled by an ice-water bath. The ice-water cooled crude solution was centrifuged at 6500 RPM for 10 min. After the centrifugation, the supernatant was discarded and the particles were re-dispersed in toluene. In order to obtain narrow size distribution of NCs, the solution was again centrifuged at 2500 RPM for 5 min. After this centrifugation, the supernatant was collected.

We calculated the quantum dot concentration in the NC film drop-cast on the silicon photonic crystal to about 5 to 6·10$^{12}$ cm$^{-2}$, considering the concentration in the initial solution (0.35 µmol/L), the volume of one droplet for drop-casting (14 µL; diameter of 3 mm) and the area of the nearly circular drop-cast spot on the sample (0.5 cm$^2$; diameter of 8 mm). As cross check, we calculated the amount of NCs in one monolayer to about 8·10$^{11}$ cm$^{-2}$ (considering the NC size of 9.4 nm plus 2 nm capping). The 80 nm thick NC film hence consists of about 7 monolayers, yielding a NC concentration of about 5 to 6·10$^{12}$ cm$^{-2}$, which is consistent with the value above.

*Fabrication of silicon photonic crystal slabs*

We fabricate crystalline silicon photonic crystal slabs in hexagonal nanohole geometry on glass substrates on an initial area of 5 x 5 cm$^2$ by a process sequence comprising nanoimprint-lithography as well as silicon thin-film deposition and crystallization techniques. Starting point is a nanoimprint template written into a silicon wafer with hexagonally arranged nano-columns exhibiting a period of 600 nm, a column diameter of 300 nm and a column height of 500 nm by Eulitha AG, Switzerland, serving as master structure. A soft nanoimprint stamp is prepared as mold for the replication process by pouring poly-(dimethyl) siloxane (PDMS) onto the template and subsequent curing at 70°C. We replicate the columnar template structure by imprinting this PDMS-stamp onto glasses coated by an organic-inorganic sol-gel consisting of a mixture of methyl-tri-methoxy-silane and tetra-methoxy-ortho-silicate [41]. The following broadband UV-curing for 5 minutes and – after having peeled off the PDMS soft stamp - thermal annealing for 60 minutes at 600°C removes organic residues and yields a silica replica of the original template considering a shrinkage of the columnar features of about 50%. The resulting hexagonally arranged silica nano-columns on glass substrates exhibit a period of 600 nm and a column height of 230 nm. We use these high-temperature stable nanostructured sol-gel coated glasses as substrate for the fabrication of crystalline silicon photonic crystal slabs in nanohole geometry. First, an amorphous silicon layer of around 120 nm is deposited by electron-beam evaporation, a directional non-conformal deposition method, at a heater temperature of around 450°C. A self-organized solid phase crystallization process by thermal annealing for 20 hours at 600°C yields a silicon layer with partially crystalline and partially amorphous regions. While the crystallization of the silicon deposited on steep flanks of the columnar features is suppressed, the bulk part of the layer becomes crystalline. The residual amorphous silicon parts surrounding the columnar silica features of the substrate are selectively removed by using a wet-chemical etch solution consisting of a phosphorous acid/nitric acid/hydrofluoric acid based silicon etchant for 2-4 seconds. In this intermediate stage, the structure consists of a crystalline silicon film with hexagonally arranged nanoholes. From these nanoholes silica pillars covered by a crystalline silicon cap protrude as the thickness of the deposited silicon film is smaller than the height of the silica nano-columns. In the last step, the protruding silica-silicon pillars are removed by mechanical abrasion using a fuzz-free tissue. The final result is a 5 x 5 cm$^2$ large glass substrate covered by a crystalline silicon photonic crystal slab in hexagonal nanohole array geometry with 115 nm layer thickness, 600 nm period and 365 nm hole diameter.

*Angular resolved reflectance measurements*

Angular resolved reflectance was measured using a Lambda 1050 UV/Vis Spectrophotometer by Perkin Elmer with automated reflectance/transmittance analyzer (ARTA) supplement [42]. The sample is mounted on a rotatable holder with the rotation axis exactly on the sample surface. When the sample is rotated, i.e. when angle of incidence is varied, the detector movement is coordinated such that the detector position always corresponds to the specularly reflected beam. Rotation angles close to normal incidence (-10° < ϑ < +10°) are omitted, because the detector cannot be moved directly in front of the light entrance port in this case.

*Excitation fluence determination*

For determination of the intensity impinging on the sample surface at normal incidence we considered a circular Gaussian beam profile with diameter $FWHM = (500 \pm 10)$ µm measured using a profile sensor (S9132, Hamamatsu) made by Pascher Instruments AB, a pulse length $\tau_p = (80 \pm 10)$ fs and a temporal Gaussian pulse shape, a repetition rate of $\nu_{rep} = 1$ kHz and a mean beam power $P_{mean}$ measured using a thermal effect-based power meter (Model 1918-C, Newport). We calculated the peak power of the optical pulse $P_{peak}$ by $P_{peak} = 0.94 \cdot P_{mean}/(\nu_{rep} \cdot \tau_p)$, as well as the peak intensity $I_{peak}$ by $I_{peak} = P_{peak}/(\pi w^2/2)$ with $w = 0.8495 \cdot FWHM$.

*Numerical simulations*

For computing reflectance spectra and field energy enhancement we model the structure as a periodic array, with a hexagonal unit cell as depicted in Fig. 2d. In accordance with the experimental setup, the periodicity length is 600 nm, the height of the Si layer is 116 nm, the diameter of the pores in the Si layer is 367 nm, and the height of the conformal QD layer is 80 nm. For $SiO_2$, a refractive index of n = 1.53, and for Si, tabulated data (E.D. Palik) is used. The time-harmonic Maxwell equations are solved using an adaptive finite-element method implemented in the commercial solver JCMsuite [29]. The unit cell of the periodic array is meshed with about 1600 prismatoidal mesh elements with sidelengths between 25 nm and 70 nm. Second order finite-elements [29] are used for the FEM discretization of the computed fields. With this numerical setting, the computation time on a standard desktop computer for a single, time-harmonic simulation is about 15 seconds. We have checked numerical accuracy by both, refining the spatial meshing as well as increasing the polynomial order of the FEM discretization in convergence test simulations. In order to obtain angle- and wavelength-dependent spectra, the wavelength and angle of incidence of the plane-wave excitation is varied in a series of computations. Reflectance and field energy enhancement are obtained from the near fields using field-integration post-processes.

## ASSOCIATED CONTENT

The manuscript is accompanied by **Supporting Information** containing
- Scanning electron microscopic image of the $CsPbBr_3$ coating on bulk and nanotextured parts of the silicon layer.
- Simulated angular resolved reflectance data for a wide range of effective refractive indices of the quantum dot containing effective medium.
- Simulated electric field energy enhancement in the 80 nm thick $CrPbBr_3$ quantum dot containing layer at short wavelengths (500 nm)
- Photonic crystal design: Simulated angular resolved reflectance data for a wide range of thicknesses of the silicon photonic crystal slab
- Fitting details for the determination of the upper intensity threshold
- Details on the $(\cos\vartheta)^2$–characteristic of photoluminescence signal dependent on the incident angle $\vartheta$

## AUTHOR INFORMATION

Corresponding Author

E-mail: christiane.becker@helmholtz-berlin.de.

Author Contributions

The manuscript was written through contributions of all authors. All authors have given approval to the final version of the manuscript.

Funding Sources

German Federal Ministry of Education and Research (BMBF)

Deutsche Forschungsgemeinschaft (DFG)

Einstein Foundation Berlin

Swedish Research Council

Knut and Alice Wallenberg Foundation

Swedish Energy Agency

NanoLund

Crafoord Foundation

ACKNOWLEDGMENT

We thank Carola Klimm from Helmholtz-Zentrum Berlin for SEM imaging. We acknowledge the German Federal Ministry of Education and Research for funding the research activities of the Nano-SIPPE group within the program NanoMatFutur (No. 03X5520). The simulation results were obtained at the Berlin Joint Lab on Optical Simulations for Energy Research (BerOSE) of Helmholtz-Zentrum Berlin für Materialien und Energie, Zuse Institute Berlin and Freie Universität Berlin. Work in Lund was financed by the Swedish Research Council (VR), the Knut and Alice Wallenberg Foundation, the Swedish Energy Agency, NanoLund and the Crafoord Foundation. We acknowledge Einstein Foundation Berlin for funding within ECMath-OT9 and Deutsche Forschungsgemeinschaft (DFG) for funding within SFB787-B4.